\documentclass[nofootinbib,twocolumn,preprintnumbers]{revtex4-1}

\usepackage{amsmath,amssymb,graphicx,booktabs,bm,psfrag,color,euscript}
\usepackage[utf8]{inputenc}

\newcommand{\A}{{\EuScript A}}
\newcommand{\scetbsm}{SCET$_\mathrm{BSM}$}
\newcommand{\nsl}{\rlap{\hspace{0.25mm}/}{n}}
\newcommand{\delsl}{\rlap{\hspace{0.25mm}/}{\partial}}
\newcommand{\vsl}{\rlap{\hspace{0.25mm}/}{v}}
\newcommand{\Asl}{\rlap{\hspace{0.6mm}/}{\A}}
\newcommand{\Dsl}{\rlap{\hspace{0.75mm}/}{D}}

\begin{document}

\preprint{MITP/19-007}
\preprint{ZU-TH~06/19}

\title{Effective Theory for a Heavy Scalar Coupled to the SM via Vector-Like Quarks}

\author{Stefan Alte$^a$}
\author{Matthias K\"onig$^b$}
\author{Matthias Neubert$^{a,c}$}

\affiliation{${}^a$PRISMA Cluster of Excellence {\em\&} MITP, Johannes Gutenberg University, 55099 Mainz, Germany\\
${}^b$Physik-Institut, Universit\"at Z\"urich, CH-8057, Switzerland \\ 
${}^c$Department of Physics {\em\&} LEPP, Cornell University, Ithaca, NY 14853, U.S.A.}

\date{February 12, 2019}

\begin{abstract}
\vspace{2mm}
We illustrate the application of the recently developed \scetbsm{} framework in the context of a specific model, in which the Standard Model (SM) is supplemented by a heavy scalar $S$ and three generations of heavy, vector-like quarks $\Psi$. We construct the appropriate effective field theory for two-body decays of $S$ into SM particles. We explicitly compute the Wilson coefficients of the \scetbsm{} operators appearing at leading and next-to-leading order (NLO) in an expansion in powers of $v/M_S$, as well as for a subset of operators arising at NNLO, retaining the full dependence on the ratio $M_S/M_\Psi$. For the phenomenologically most relevant decay channels of the heavy scalar, we study the impact of resummation effects of Sudakov logarithms on the decay rates.
\end{abstract}

\maketitle

\section{Introduction}

Following the discovery of a new particle with a mass far above the electroweak scale $v\approx 246$\,GeV, a program for studying its couplings to the Standard Model (SM) would be of highest priority. In the likely situation where the new resonance is the first member of a richer sector of new physics, the appropriate way to study its decay and production processes must rely on an effective field theory (EFT) framework. The main reason is that other, yet undiscovered heavy particles can couple to both the SM and the new resonance $S$ and hence affect its interactions. Secondly, the large scale hierarchy between the mass of the heavy resonance and the weak scale, which (roughly) sets the masses of the SM particles, introduces large Sudakov double logarithms in the calculation of decay rates and production cross sections, which must be resummed to all orders of perturbation theory. Finally, for the most interesting case where the mass of the new resonance is close to the masses $M_i$ of yet undiscovered states, there is short-distance physics associated with both scales, which must be disentangled from the longer-distance physics associated with the electroweak scale. 

We have shown in \cite{Alte:2018nbn} that the appropriate EFT to deal with this scenario must be based on an effective Lagrangian built out of non-local light-ray operators defined in soft-collinear effective theory (SCET) \cite{Bauer:2000yr,Bauer:2001ct,Bauer:2001yt,Beneke:2002ph}. Our theory called \scetbsm{} provides a systematic expansion of the decay amplitudes of the new heavy particle in powers of $\lambda=v/M_S\ll 1$. For the case of a scalar resonance $S$ transforming as a singlet under the SM gauge group, we have constructed the complete operator basis at leading and subleading order in the expansion, corresponding to operators of ${\mathcal O}(\lambda^2)$ and ${\mathcal O}(\lambda^3)$, respectively.

The leading-order effective Lagrangian for two-body decays of $S$ consists of operators in which $S$ is coupled to two effective bosonic fields, which describe so-called collinear particles moving along directions $\bm{n}_1$ and $\bm{n}_2$, which point back-to-back in the rest frame of the decaying resonance. One has \cite{Alte:2018nbn}
\begin{equation}\label{Leff2}
\begin{aligned}
   {\mathcal L}_{\rm eff}^{(2)} 
   &= M\,C_{\phi\phi}(M,M_S)\,O_{\phi\phi} \\
   &\mbox{}+ M \sum_A \Big[ C_{AA}(M,M_S)\,O_{AA} + \widetilde C_{AA}(M,M_S)\,\widetilde O_{AA} \Big] \,.
\end{aligned}
\end{equation} 
Here $M$ denotes the characteristic mass scale of unresolved new heavy particles. The sum extends over the three gauge groups of the SM: $A=B$ for $U(1)_Y$, $A=W$ for $SU(2)_L$, and $A=G$ for $SU(3)_c$. The relevant \scetbsm{} operators have the form (a summation over the group index $a$ is understood for non-abelian fields)
\begin{equation}\label{eq2}
\begin{aligned}
   O_{\phi\phi} &= S_v\,\big( \Phi_{n_1}^\dagger \Phi_{n_2} + \Phi_{n_2}^\dagger \Phi_{n_1} \big) \,, \\
   O_{AA} &= S_v\,g_{\mu\nu}^\perp \,\A_{n_1}^{\mu,a}\,\A_{n_2}^{\nu,a} \,, \\
   \widetilde O_{AA} &= S_v\,\epsilon_{\mu\nu}^\perp\,\A_{n_1}^{\mu,a}\,\A_{n_2}^{\nu,a} \,.
\end{aligned}
\end{equation} 
Here $S_v$ is an effective field for the heavy resonance defined as in heavy-quark effective theory \cite{Eichten:1989zv,Georgi:1990um,Falk:1990pz,Neubert:1993mb}, with $v$ denoting its 4-velocity. The reference vectors $n_1$ and $n_2$ indicate the directions of large momentum flow of the final-state particles. The effective fields consist of so-called ``gauge covariant building blocks'' \cite{Bauer:2002nz,Hill:2002vw} $\Phi$ and $\A$ containing the Higgs doublet and the transversely polarized gauge fields, respectively, dressed up with Wilson lines in the appropriate representation of the gauge group. The Lorentz indices of the gauge fields can be contracted with either the symmetric tensor $g_{\mu\nu}^\perp$ or the antisymmetric tensor $\epsilon_{\mu\nu}^\perp$ defined in the plane orthogonal to $n_1$ and $n_2$. Note that the different fields in the operators in (\ref{eq2}) interact only via soft quanta, since there is only a single collinear field in each sector; hard interactions with virtualities of order $M_S^2$ or $M^2$ are integrated out in the construction of the effective Lagrangian and are contained in the Wilson coefficient functions.

Note the important fact that the Wilson coefficients in (\ref{Leff2}) depend on both, the mass $M_S$ of the scalar resonance and the parameter $M$ representing the typical mass scale of other, yet undiscovered heavy particles. As we shall see below, in this way our effective theory sums infinite towers of local operators in the conventional EFT approach. In some sense, the Wilson coefficients in our Lagrangian can be regarded as form factors depending on the large momentum transfers $q^2={\mathcal O}(M_S^2)$ flowing through Feynman diagrams, which can resolve the small non-localities corresponding to exchanges of the heavy VLQs. 

At subleading order in power counting the operator basis contains five different types of operators, all of which consist of fermion bilinears along with a Higgs doublet or a gauge field, see Section~\ref{sec:treematch}. In Section~\ref{sec:lam4} we study some aspects of the extension of the effective Lagrangian to ${\mathcal O}(\lambda^4)$, which is necessary to describe the two-body decay $S\to Zh$.

In this work, we illustrate the \scetbsm{} approach by considering a concrete extension of the SM featuring a heavy, gauge-singlet scalar field $S$ along with three generations of heavy, vector-like quarks. Vector-like fermions play an important role in models of partial compositeness \cite{Kaplan:1991dc}, as realized e.g.\ in composite-Higgs models (see e.g.\ \cite{Dugan:1984hq,Agashe:2004rs,Contino:2010rs}) and scenarios featuring a warped extra dimension \cite{Randall:1999ee,Grossman:1999ra,Gherghetta:2000qt}. Extensions of the SM featuring both vector-like fermions and a singlet scalar are among the popular simplified models for dark matter (see e.g.\ \cite{Buckley:2014fba,Harris:2014hga}).

\section{High-energy extension of the SM}

The benchmark model we explore in this paper is an extension of the SM by a real scalar $S$, transforming as a singlet under the SM gauge group, and (three generations of) a vector-like quark (VLQ) doublet $\Psi$, transforming as $(\bm{3},\bm{2})_{1/6}$. Besides the Higgs portal, the VLQs mediate the renormalizable interactions between the SM and the new sector. We assume that the mass of the scalar and the masses of the VLQs are both much heavier than the electroweak scale $v\approx 246$\,GeV. The most general Lagrangian of our model is
\begin{equation}\label{Lfull}
\begin{aligned}
   {\mathcal L}_\mathrm{UV} 
   &= {\mathcal L}_\mathrm{SM} 
    + \frac{1}{2}(\partial_\mu S)(\partial^\mu S) - \frac{M_S^2}{2}\,S^2 - \frac{\lambda_3}{3!}\,S^3
    - \frac{\lambda_4}{4!}\,S^4 \\[1mm]
   &\quad\mbox{}+ \bar\Psi (i\Dsl-\bm{M}) \Psi - \big( \bar\Psi \tilde\phi\,\bm{G}_u u_R + \bar\Psi \phi\,\bm{G}_d d_R 
    + \mathrm{h.c.} \big) \\
   &\quad\mbox{} 
    - \kappa_1 S\,\phi^\dagger\phi - \frac{\kappa_2}{2}\,S^2 \phi^\dagger\phi \\
   &\quad\mbox{}- S\,\bar\Psi \big(\bm{X} - i\gamma_5\tilde{\bm{X}}\big) \Psi 
    - S\,\big( \bar\Psi\,\bm{V}_Q\, Q_L + \mathrm{h.c.} \big) \,.
\end{aligned}
\end{equation}
The second line contains the couplings of the VLQs to SM fields, where $\tilde\phi_a=\epsilon_{ab}\phi^\ast_b$. There is no need to include the gauge-invariant terms $\bar\Psi i\gamma_5\tilde{\bm{M}}\Psi-(\bar\Psi\,\bm{G}_Q Q_L+\text{h.c.})$, since they can be removed by unitary transformations of the quark fields. The terms in the third line contain the portal couplings of the heavy scalar to the Higgs field. Note that the couplings $\kappa_1$ and $\lambda_3$ have mass dimension~1. The interactions in the last line describe the couplings of $S$ to the VLQs and SM quarks. We assume that the parameters $\lambda_i$ in the scalar potential are chosen such that the scalar field $S$ does not acquire a vacuum expectation value. For the same reason, we have omitted the tadpole term $\lambda_1 S$ from the potential. 

Boldface symbols in (\ref{Lfull}) denote matrices in generation space. The matrices $\bm{G}_{u,d}$ and $\bm{V}_Q$ are arbitrary complex matrices, while $\bm{M}$, $\bm{X}$ and $\tilde{\bm{X}}$ are hermitian. Without loss of generality we work in the mass basis for the VLQs, where $\bm{M}$ is a real, positive diagonal matrix. For simplicity, we assume that the three mass eigenvalues are degenerate, i.e.\ $\bm{M}=M\,\bm{1}$. The common mass of the VLQs is then identified with the ``new physics scale'' $M$ in (\ref{Leff2}). 

Suppose that the heavy scalar $S$ has been discovered, while the VLQs have not yet been observed experimentally. Our goal is to construct an EFT describing the interactions of $S$ with SM particles. The appropriate EFT in such a scenario is the \scetbsm{} \cite{Alte:2018nbn}. It would be straightforward to extend our analysis to the case of vector-like fermions with different quantum numbers. However, in order to keep the presentation as transparent as possible, we find it advantageous to consider the simplest case of a single type of VLQ.

\section{Tree-level matching onto \scetbsm}
\label{sec:treematch}

When the full theory in (\ref{Lfull}) is matched onto the \scetbsm{} two types of short-distance modes are integrated out: First, one removes virtual exchanges of the VLQs, which do not appear as external states in the EFT (since these particles are assumed to be yet undiscovered). In addition, one integrates out off-shell fluctuations of the SM fields as well as of the scalar field $S$ carrying virtualities of order $q^2\sim M_S^2$. While the first step is standard, the second step differentiates the \scetbsm{} approach from local EFTs such as the SMEFT \cite{Weinberg:1979sa,Wilczek:1979hc,Buchmuller:1985jz,Leung:1984ni,Grzadkowski:2010es}. 

\subsection{Integrating out the vector-like quarks}

At tree level, the heavy VLQs can be integrated out by solving their classical equations of motion. This yields the ``non-local effective Lagrangian''
\begin{equation}\label{Leff1}
\begin{aligned}
   {\mathcal L}_\mathrm{eff} 
   &= {\mathcal L}_\mathrm{SM} + \frac{1}{2}(\partial_\mu S)(\partial^\mu S) 
    - \frac{M_S^2}{2}\,S^2 - \frac{\lambda_3}{3!}\,S^3 - \frac{\lambda_4}{4!}\,S^4 \\
   &\quad\mbox{}- \kappa_1 S\,\phi^\dagger\phi - \frac{\kappa_2}{2} S^2 \phi^\dagger\phi \\
   &\quad\mbox{}- \bar F\,\frac{1}{i\Dsl - M - S \big(\bm{X} - i\gamma_5\tilde{\bm{X}}\big)}\,F \,,    
\end{aligned}
\end{equation}
where
\begin{equation}
   F = \tilde\phi\,\bm{G}_u u_R + \phi\,\bm{G}_d d_R + S\,\bm{V}_Q Q_L \,.
\end{equation}
Note that the heavy scalar field $S$ is still a propagating field at this stage, and indeed the last term in (\ref{Leff1}) contains couplings of SM fields to an arbitrary number of $S$ fields. The terms of zeroth order in $S$ read
\begin{equation}
   {\mathcal L}_\mathrm{eff} \big|_{S^0}
   = {\mathcal L}_\mathrm{SM} + \bar F_0\,\frac{1}{M-i\Dsl}\,F_0 \,,
\end{equation}
where $F_0=\tilde\phi\,\bm{G}_u u_R+\phi\,\bm{G}_d d_R$. Expanding the denominator in powers of covariant derivatives would generate an infinite set of higher-dimensional, gauge-invariant operators, which account for the virtual effects of heavy VLQs on the interactions among SM particles in the context of the SMEFT. 

For our purposes the most relevant terms in (\ref{LlinearS}) are those linear in $S$. They are
\begin{equation}\label{LlinearS}
\begin{aligned}
   {\mathcal L}_\mathrm{eff} \big|_{S^1}
   &= - \kappa_1 S\,\phi^\dagger\phi + \left[ S\,\bar Q_L \bm{V}_Q^\dagger\,\frac{1}{M-i\Dsl}\,F_0 + \text{h.c.} \right] \\
   &\quad\mbox{}- \bar F_0\,\frac{1}{M-i\Dsl}\,S \big(\bm{X} - i\gamma_5\tilde{\bm{X}}\big)\, 
    \frac{1}{M-i\Dsl}\,F_0 \,.
\end{aligned}
\end{equation}
In order to match this expression onto the \scetbsm{} effective Lagrangian describing two-body decays of the heavy scalar $S$, we replace the SM fields by fields in the EFT. The relevant fields are the soft field $S_v$ for the heavy resonance and collinear fields describing particle jets moving along light-like directions $n_1^\mu=(1,\bm{n}_1)$ and $n_2^\mu=(1,\bm{n}_2)$. The precise definitions of these fields, which include collinear Wilson lines, can be found in \cite{Alte:2018nbn}. For the special case of the Higgs doublet, the low-energy theory also contains a soft field $\Phi_0$ carrying no 4-momentum. After electroweak symmetry breaking this field is set to the Higgs vacuum expectation value. The relevant replacement rules are extremely simple:
\begin{equation}\label{fieldmatch}
\begin{aligned}
   \phi &\to \Phi_0 + \Phi_{n_1} + \Phi_{n_2} + \dots \,, \\[1mm]
   \psi &\to \psi_{n_1} + \psi_{n_2} + \dots \,, \\[1mm]
   g A^{\mu,a} &\to \A_{n_1}^{\mu,a} + \A_{n_2}^{\mu,a} + \dots \,.
\end{aligned}
\end{equation}
Here $\psi=Q_L,u_R,d_R$ denotes a generic SM quark field, while $A=B,W,G$ is a generic gauge field. The effective gauge fields in \scetbsm{} include the gauge couplings in their definition. The collinear quark and gauge fields are subject to the constraints $\nsl_i\,\psi_{n_i}=0$ and $\bar n_i\cdot\A_{n_i}^a=0$, where $\bar n_i^\mu=(1,-\bm{n}_i)$. Note that the components $\bar n_i\cdot A^a$ of the gauge fields are contained in the Wilson lines of the effective theory. The collinear fields satisfy simple power counting rules in the expansion parameter $\lambda=v/M_S$ of \scetbsm: the fields $\Phi_0$, $\Phi_{n_i}$, $\psi_{n_i}$ and $\A_{\perp,n_i}^{\mu,a}$ are all of ${\mathcal O}(\lambda)$, whereas the longitudinal gauge fields $n_i\cdot\A_{\perp,n_i}^a$ are of ${\mathcal O}(\lambda^2)$. The subscript $\perp$ refers to the components of an $n_i$-collinear gauge field perpendicular to the 4-vectors $n_i^\mu$ and $\bar n_i^\mu$. Derivatives acting on $n_i$-collinear fields can be decomposed into the components $\bar n_i\cdot\partial={\mathcal O}(\lambda^0)$, $\partial_\perp^\mu={\mathcal O}(\lambda)$ and $n_i\cdot\partial={\mathcal O}(\lambda^2)$. The dots in (\ref{fieldmatch}) stand for soft fields, which are power-suppressed relative to the collinear fields and will play no role for our discussion. 

\begin{figure}
\begin{center}
\includegraphics[scale=0.6]{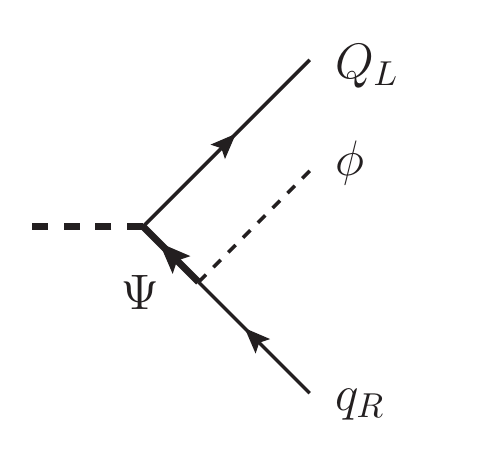}
\includegraphics[scale=0.6]{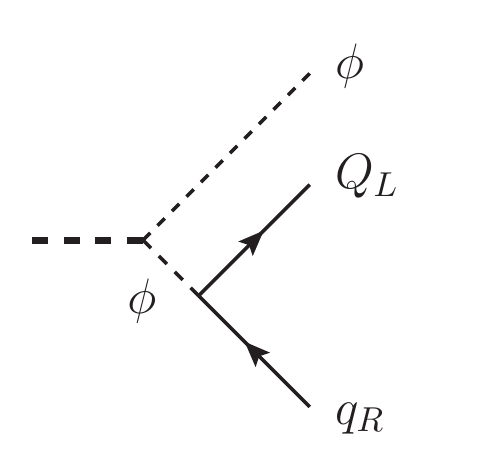}
\vspace{-5mm}
\end{center}
\caption{\label{fig:step1}
Tree-level diagrams giving rise to the effective Lagrangians (\ref{res1}) [left] and (\ref{res2}) [right]. Thick lines denote $S$ and the VLQs, whereas the thin lines represent SM particles.}
\end{figure}

It is now straightforward to extract from (\ref{LlinearS}) the terms of leading order in the $\lambda$ expansion. Obviously, the first term on the right-hand side generates the tree-level contribution
\begin{equation}\label{Cphiphitree}
   C_{\phi\phi} = - \frac{\kappa_1}{M}
\end{equation} 
to the Wilson coefficient of the scalar operator $O_{\phi\phi}$ in the \scetbsm{} Lagrangian (\ref{Leff2}). After the introduction of SCET fields the quantity $F_0$ is of ${\mathcal O}(\lambda^2)$, while $Q_L$ is of ${\mathcal O}(\lambda)$. Hence the leading terms in the Lagrangian originating from VLQ exchange are of ${\mathcal O}(\lambda^3)$ and arise from the term in brackets in the first line of (\ref{LlinearS}). Since gauge fields in \scetbsm{} are always power suppressed, we can expand the inverse derivative operator sandwiched between spinor fields of opposite chirality in the form
\begin{equation}
   \frac{1}{M-i\Dsl} 
   \to \frac{M}{M^2+\Box} + {\mathcal O}(\lambda) \,.
\end{equation}
The Laplace operator in the denominators of these expressions must only be kept if the product of fields on which this operator acts has virtuality of order $M_S^2$. We thus obtain
\begin{equation}\label{res1}
\begin{aligned}
   {\mathcal L}_\mathrm{eff} \big|_{S^1}^{\lambda^3}
   &= \frac{1}{M} \sum_{q=u,d} \bigg[ 
    S_v\,\bar Q_{L,n_1} \bm{V}_Q^\dagger\bm{G}_q \big( \Phi_0 + \Phi_{n_2} \big) q_{R,n_2} \\   
   &\quad\mbox{}+ S_v\,\bar Q_{L,n_1} \bm{V}_Q^\dagger\bm{G}_q\,\frac{M^2}{M^2+\Box}\,\Phi_{n_1} q_{R,n_2}
    + \mathrm{h.c.} \bigg] \\
   &\quad\mbox{}+ (n_1\leftrightarrow n_2) \,,
\end{aligned}
\end{equation}
where for $q=u$ the doublet $\Phi$ must be replaced by $\tilde\Phi$. The first graph in Figure~\ref{fig:step1} shows a diagram in the complete theory giving rise to these matching contributions.

\subsection{Integrating out off-shell fluctuations}

If the portal coupling $\kappa_1$ in (\ref{Lfull}) is non-zero, then the second diagram shown in Figure~\ref{fig:step1} produces another tree-level matching contribution, in which the propagator for the Higgs doublet carries a virtuality of order $q^2\sim M_S^2$. The corresponding contribution to the effective Lagrangian can be written in the form
\begin{equation}\label{res2}
\begin{aligned}
   \Delta{\mathcal L}_\mathrm{eff} \big|_{S^1}^{\lambda^3}
   &= \sum_{q=u,d} \kappa_1 S_v \big( \Phi_0^a + \Phi_{n_1}^a + \Phi_{n_2}^a \big) 
    \frac{1}{\Box}\,\bar Q_{L,n_1}^a \bm{Y}_q\,q_{R,n_2} \\
   &\quad\mbox{}+ \mathrm{h.c.} + (n_1\leftrightarrow n_2) \,,
\end{aligned}
\end{equation}
where the inverse Laplace operator arises from the Higgs propagator. The sum of (\ref{res1}) and (\ref{res2}) gives the complete tree-level effective Lagrangian at ${\mathcal O}(\lambda^3)$.

\subsection{Wilson coefficients}

The complete basis of \scetbsm{} operators at ${\mathcal O}(\lambda^3)$ has been constructed in \cite{Alte:2018nbn}. The effective Lagrangian at this order can be written in the form (summed over $i,j$)
\begin{equation}\label{Leff3}
\begin{aligned}
   {\mathcal L}_{\rm eff}^{(3)}
   &= \frac{1}{M} \sum_{q=u,d} \bigg[ C_{Q_L\bar q_R}^{\,ij}(M,M_S)\,O_{Q_L\bar q_R}^{\,ij} \\
   &\quad\mbox{}+ \!\sum_{k=1,2} \int_0^1\!du\,C_{Q_L\bar q_R\,\phi}^{(k)\,ij}(u,M,M_S)\,
    O_{Q_L\bar q_R\,\phi}^{(k)\,ij}(u) + \mbox{h.c.} \bigg] \\
   &\quad\mbox{}+ \frac{1}{M} \sum_A \bigg[ 
    \int_0^1\!du\,C_{Q_L\bar Q_L A}^{\,ij}(u,M,M_S)\,O_{Q_L\bar Q_L A}^{\,ij}(u) \\
   &\hspace{21mm}\mbox{}+ (Q_L\to q_R) + \mbox{h.c.} \bigg] \,,
\end{aligned}
\end{equation} 
where the sum in the last lines runs over the three gauge fields $A=B,W,G$. For simplicity we consider operators containing quark fields only. We have defined the mixed-chirality operators
\begin{equation}
\begin{aligned}
   O_{Q_L\bar q_R}^{\,ij}
   &= S_v\,\bar Q_{L,n_1}^{\,i} \Phi_0\,q_{R,n_2}^{\,j} + (n_1\leftrightarrow n_2) \,, \\
   O_{Q_L\bar q_R\,\phi}^{(1)\,ij}(u)
   &= S_v\,\bar Q_{L,n_1}^{\,i} \Phi_{n_1}^{(u)}\,q_{R,n_2}^{\,j} + (n_1\leftrightarrow n_2) \,, \\
   O_{Q_L\bar q_R\,\phi}^{(2)\,ij}(u)
   &= S_v\,\bar Q_{L,n_1}^{\,i} \Phi_{n_2}^{(u)}\,q_{R,n_2}^{\,j} + (n_1\leftrightarrow n_2) \,,
\end{aligned}
\end{equation} 
and the same-chirality operators
\begin{equation}
\begin{aligned}
   O_{Q_L\bar Q_L A}^{\,ij}(u)
   &= S_v\,\bar Q_{L,n_1}^{\,i}\,\Asl_{n_1}^{\perp(u)}\,Q_{L,n_2}^{\,j}
    + (n_1\leftrightarrow n_2) \,, \\
   O_{q_R\bar q_R A}^{\,ij}(u)
   &= S_v\,\bar q_{R,n_1}^{\,i}\,\Asl_{n_1}^{\perp(u)}\,q_{R,n_2}^{\,j}
    + (n_1\leftrightarrow n_2) \,,
\end{aligned}
\end{equation} 
where $i,j$ are generation indices. When an operator contains more than two collinear fields describing particles moving in the same direction, the total collinear momentum carried by this jet is split up among the fields. Our convention is that in each operator the bosonic field carries the longitudinal momentum fraction $u\in[0,1]$, while the fermionic field carries momentum fraction $(1-u)$. From (\ref{res1}) and (\ref{res2}), we obtain for the tree-level matching conditions in matrix notation (with $q=u,d$)
\begin{equation}\label{eq:twoJetLambda3}
\begin{aligned}
   \bm{C}_{Q_L\bar q_R} &= \bm{V}_Q^\dagger \bm{G}_q - \frac{\kappa_1}{M}\,\frac{\bm{Y}_q}{\xi} \,, \\[-1mm]
   \bm{C}_{Q_L\bar q_R\phi}^{(1)} &= \frac{\bm{V}_Q^\dagger \bm{G}_q}{1-\xi u-i\epsilon}
    - \frac{\kappa_1}{M}\,\frac{\bm{Y}_q}{\xi(1-u)+i\epsilon} \,, \\
   \bm{C}_{Q_L\bar q_R\phi}^{(2)} &= \bm{V}_Q^\dagger \bm{G}_q
    - \frac{\kappa_1}{M}\,\frac{\bm{Y}_q}{\xi(1-u)+i\epsilon} \,, \\[1mm]
   \bm{C}_{Q_L\bar Q_L A} &= \bm{C}_{q_R\bar q_R A} = 0 \,; \quad A=B,W,G \,, 
\end{aligned}
\end{equation}
where we have defined $\xi=M_S^2/M^2$. The $i\epsilon$ prescriptions are those from the Feynman propagators. Note that $\kappa_1$ is naturally of order $M$. The parameter $\xi$ governs the ratio of the mass of the heavy scalar resonance, which we assume has been discovered, and the mass of the VLQs, which we assume have not yet been discovered. This ratio is in principle arbitrary, but in many realistic models is expected to be of ${\mathcal O}(1)$. The fact that \scetbsm{} correctly captures the dependence on both mass parameters is a unique feature of this EFT \cite{Alte:2018nbn}.

Analogous operators containing lepton fields also exist, and indeed they can be generated at tree level in our model via the Higgs portal interaction proportional to $\kappa_1$. However, the corresponding Wilson coefficients are strongly suppressed by the leptonic Yukawa couplings. 

The coefficients in (\ref{eq:twoJetLambda3}) are given in the weak basis. After electroweak symmetry breaking, these coefficients should be transformed to the mass basis of the SM quarks. This transformation diagonalizes the Yukawa matrices $\bm{Y}_q$, while $\bm{V}_Q^\dagger\bm{G}_q\to\bm{U}_{q_L}^\dagger\bm{V}_Q^\dagger\bm{G}_q\bm{W}_{q_R}$, where $\bm{U}_{q_L}$ and $\bm{W}_{q_R}$ with $q=u,d$ denote the rotation matrices transforming the left-handed and right-handed quark fields from the weak to the mass basis.

\section{One-loop matching}

\begin{figure}
\begin{center}
\includegraphics[scale=0.54]{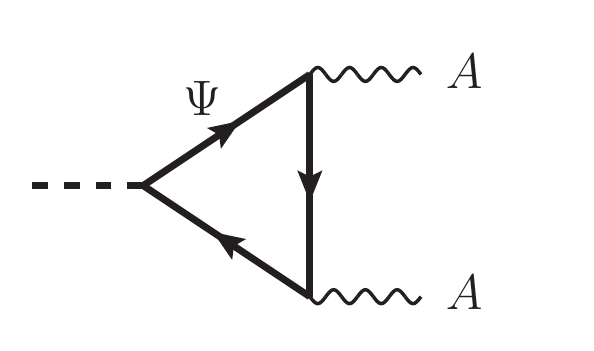} 
\hspace{-8mm}
\includegraphics[scale=0.54]{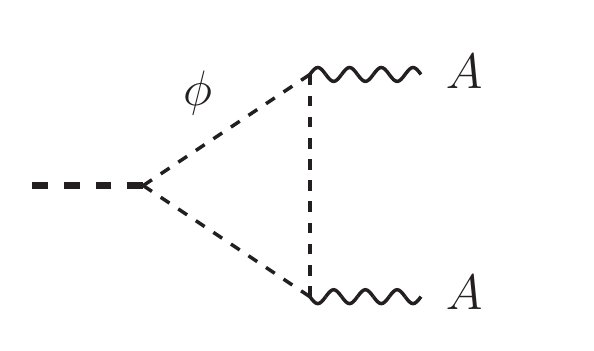}
\hspace{-8mm}
\includegraphics[scale=0.54]{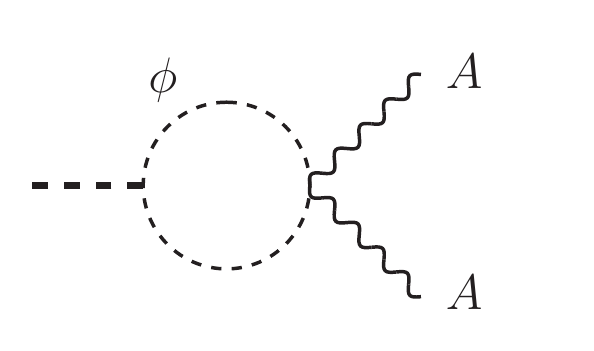}
\vspace{-8mm}
\end{center}
\caption{\label{fig:diagrams} 
One-loop diagrams contributing to the Wilson coefficients $C_{AA}$ and $\widetilde C_{AA}$ in (\ref{Leff2}). We do not show crossed graphs, in which the two boson lines are exchanged. Loops graphs involving SM fermions do not arise at leading order in $\lambda$.}
\end{figure}

With the exception of $O_{\phi\phi}$, the bosonic operators in the \scetbsm{} Lagrangian receive matching corrections starting at one-loop order. We now discuss the calculation of these corrections for the Wilson coefficients of the leading operators of ${\mathcal O}(\lambda^2)$ in (\ref{Leff2}). The relevant Feynman diagrams are shown in Figure~\ref{fig:diagrams}. The first graph contains a loop of VLQs, while the remaining diagrams feature loops with off-shell Higgs doublets. To perform the matching in the simplest possible way, we calculate these diagrams setting all SM masses to zero. Then loop graphs in the EFT are scaleless and vanish, and hence the Wilson coefficients are given directly in terms of the diagrams shown in the figure. We find (with $A=B,W,G$) 
\begin{equation}\label{eq:cAA}
\begin{aligned}
   C_{AA} &= \frac{d_A}{\pi^2}\,\mathrm{Tr}(\bm{X}) \left[ \frac{4 -\xi}{\xi}\,g^2(\xi) - 1 \right] 
    + \frac{d_A'}{4\pi^2}\,\frac{\kappa_1}{M} \,, \\
   \widetilde C_{AA} &= \frac{d_A}{\pi^2}\,\mathrm{Tr}(\tilde{\bm{X}})\,g^2(\xi) \,,
\end{aligned}
\end{equation}
where $\xi=M_S^2/M^2$ as above, and the group-theory factors $d_A$ are given by
\begin{equation}
\begin{aligned}
   d_B &= N_c Y_\psi^2 = \frac{1}{12} \,, &
   d_W &= \frac{T_F N_c}{2} = \frac34 \,, &
   d_G &= T_F = \frac12 \,, \\
   d_B' &= Y_\phi^2 = \frac14 \,, &
   d_W' &= \frac{T_F}{2} = \frac14 \,, &
   d_G' &= 0 \,.
\end{aligned}
\end{equation}
The relevant loop function reads
\begin{equation}\label{gfun}
   g(\xi) = \left\{ \begin{array}{cc}
    \displaystyle \arcsin\frac{\sqrt{\xi}}{2} \,; & \xi\le 4 \,, \\[3mm]
    \displaystyle \frac{i}{2} \ln\frac{1+\sqrt{1-4/\xi}}{1-\sqrt{1-4/\xi}} + \frac{\pi}{2} \,; ~ & \xi\ge 4 \,.
   \end{array} \right.
\end{equation}
The $\xi$-dependent contributions arise from integrating out the VLQs, while the term proportional to the Higgs-portal coupling $\kappa_1$ is obtained by integrating out loops of virtual Higgs doublets carrying virtualities of order $M_S^2$, in analogy with the discussion in the previous section.

It is instructive to study the $\xi$-dependent terms in (\ref{eq:cAA}) in more detail. Focussing on the case of $\widetilde C_{AA}$ for concreteness, and assuming that $M_S^2<4M^2$, we can expand the Wilson coefficient in powers of the ratio $\xi=M_S^2/M^2$, finding
\begin{equation}
   \widetilde C_{AA} = \frac{d_A}{2\pi^2}\,\mathrm{Tr}(\tilde{\bm{X}}) \sum_{k=1}^\infty 
    \frac{\Gamma\big(\frac{1}{2}\big)\,\Gamma(k)}{k\,\Gamma\big(\frac{1}{2}+k\big)} 
    \left( \frac{M_S^2}{4M^2} \right)^k .
\end{equation}
The first term in the sum gives a contribution to (\ref{Leff2}) which corresponds to the local dimension-5 operator $S F_{\mu\nu} \widetilde F^{\mu\nu}$, the second term corresponds to local dimension-7 operators such as $S (\partial_\alpha F_{\mu\nu}) (\partial^\alpha\widetilde F^{\mu\nu})$ or $(\Box S) F_{\mu\nu} \widetilde F^{\mu\nu}$, the third term corresponds to local dimension-9 operators, and so on. Our \scetbsm{} approach thus sums up an infinite tower of local operators. In an extension of the SMEFT consisting of local operators built out of SM field and the field $S$ (see e.g.\ \cite{Franceschini:2015kwy,Gripaios:2016xuo}), one would typically only include the leading dimension-5 operators. In realistic scenarios where $M_S\sim M$, however, all contributions are of the same order. 

\begin{figure}
\begin{center}
\includegraphics[scale=0.54]{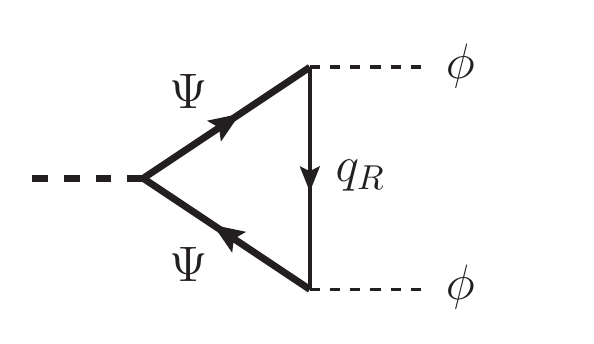}
\hspace{-3mm}
\includegraphics[scale=0.54]{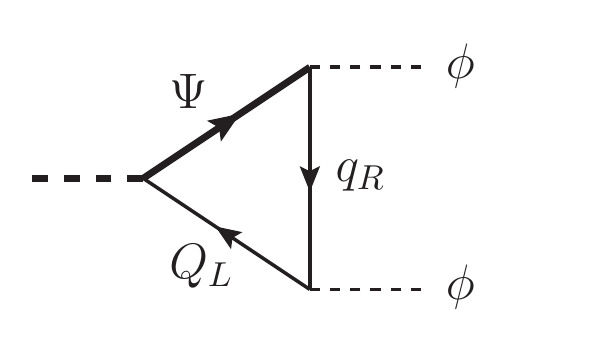}
\hspace{-3mm}
\includegraphics[scale=0.54]{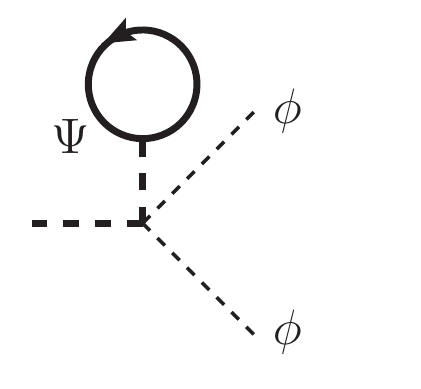}
\includegraphics[scale=0.54]{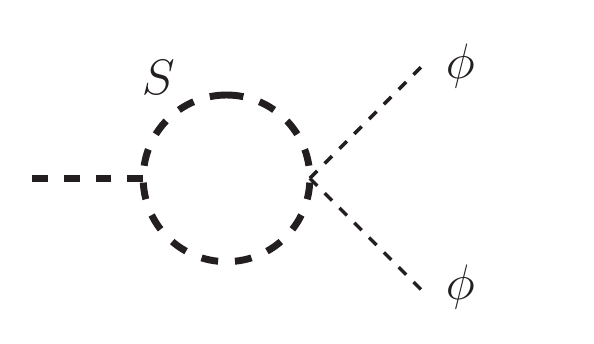}
\includegraphics[scale=0.54]{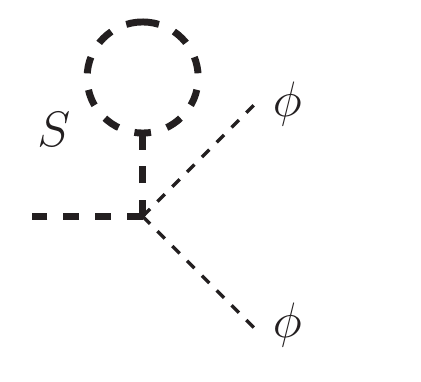}
\vspace{-5mm}
\end{center}
\caption{\label{fig:morediagrams} 
One-loop diagrams contributing to the coefficient $C_{\phi\phi}^{(1)}$ in (\ref{eq18}).}
\end{figure}

The one-loop matching calculation for the coefficient $C_{\phi\phi}$ in (\ref{Leff2}) is more involved. We write the result in the form
\begin{equation}\label{eq18}
   C_{\phi\phi} 
   = - \frac{\kappa_1}{M} \left( 1 + \delta_{\kappa_1} \right) + C_{\phi\phi}^{(1)} \,.
\end{equation}
The quantity $\delta_{\kappa_1}$ contains the loop corrections to the tree-level result in (\ref{Cphiphitree}), while $C_{\phi\phi}^{(1)}$ contains contributions to the Wilson coefficient involving couplings other than $\kappa_1$. The relevant diagrams for the latter terms are shown in Figure~\ref{fig:morediagrams}. We obtain 
\begin{equation}
\begin{aligned}
   C_{\phi\phi}^{(1)}
   &= \frac{N_c}{8\pi^2}\,\mathrm{Tr}\big[\bm{X}(\bm{G}_u\bm{G}_u^\dagger+\bm{G}_d\bm{G}_d^\dagger) \big] \\
   &~~\quad\times \left[ 2\ln\frac{M^2}{\mu^2} - 3 + 2\sqrt{\frac{4-\xi}{\xi}}\,g(\xi)
    + \frac{4}{\xi}\,g^2(\xi) \right] \\
   &\quad\mbox{}+ \frac{N_c}{8\pi^2}\,
    \mathrm{Re\,Tr}\big[\bm{V}_Q(\bm{Y}_u\bm{G}_u^\dagger+\bm{Y}_d\bm{G}_d^\dagger) \big] \\
   &~~\quad\times \left[ 2\ln\frac{M^2}{\mu^2} - 3 - \frac{1-\xi}{\xi} \ln(1-\xi-i\epsilon) \right] \\
   &\quad\mbox{}- \frac{N_c\kappa_2}{2\pi^2\xi}\,
    \mathrm{Tr}(\bm{X}) \left( \ln\frac{M^2}{\mu^2} - 1 \right) \\
   &\quad\mbox{}- \frac{\kappa_2\lambda_3}{32\pi^2 M} \left( \frac{\pi}{\sqrt3} - 1 \right) .
\end{aligned}
\end{equation}
The calculation of $\delta_{\kappa_1}$ is discussed in Appendix~\ref{app:A}. Unlike the results shown in (\ref{eq:cAA}), these expressions contain an explicit dependence on the renormalization scale $\mu$, at which the operators and Wilson coefficients are defined (in the $\overline{\rm MS}$ scheme). The matching results presented here refer to a scale $\mu\sim M$, at which they do not contain any large logarithms; the evolution to lower scales will be discussed later in Section~\ref{sec:pheno}. The scale dependence of the coefficient $C_{\phi\phi}^{(1)}$ must be compensated by the scale dependence of the portal coupling $\kappa_1$ in (\ref{eq18}).

\section{\boldmath One-loop matching for $S\to Zh$}
\label{sec:lam4}

There is one potential two-body decay of a heavy scalar resonance $S$ that cannot be described using the operators arising at leading and subleading order in SCET power counting. This is the mode $S\to Zh$, where the $Z$ boson is longitudinally polarized. Only the CP-odd component of the scalar can decay to this final state, which makes this channel interesting to study the CP properties of a new scalar resonance \cite{Bauer:2016ydr,Bauer:2016zfj}. The following discussion is significantly more technical than that in the previous two sections and can be skipped in a first reading.

\begin{figure*}
\begin{center}
\includegraphics[scale=0.6]{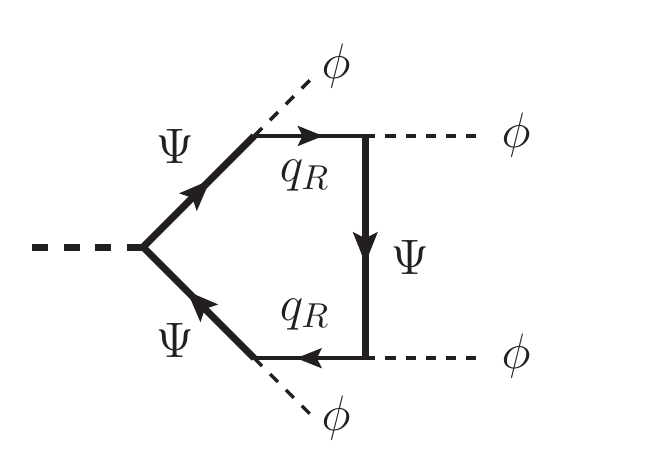}
\includegraphics[scale=0.6]{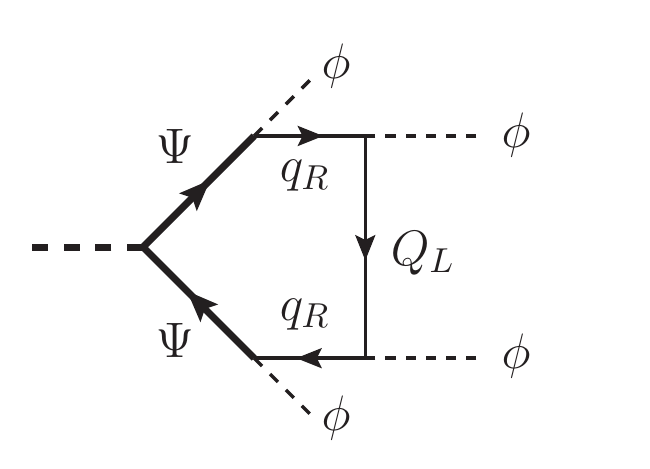}
\includegraphics[scale=0.6]{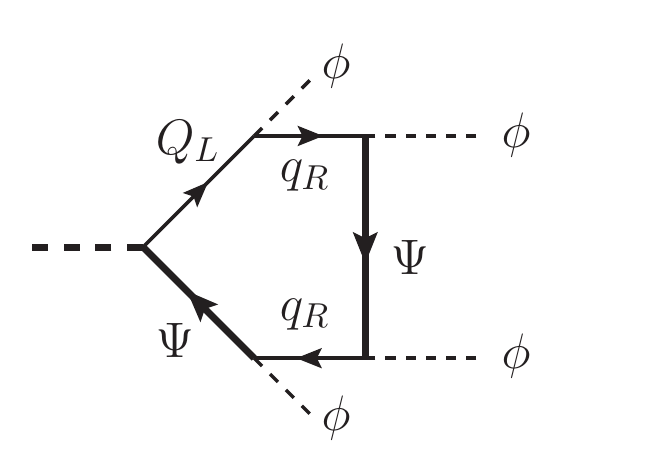}
\includegraphics[scale=0.6]{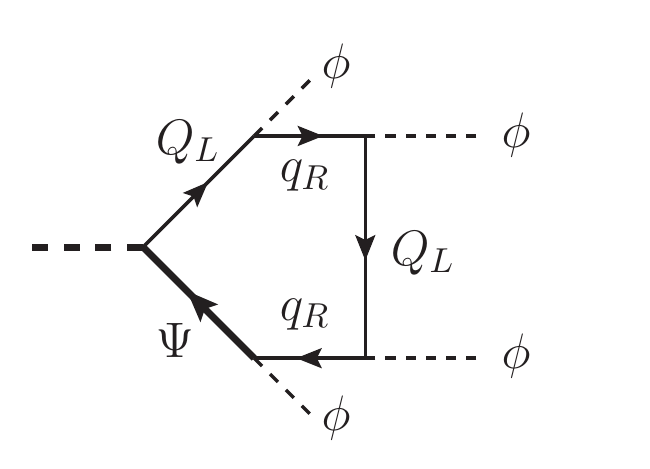}
\includegraphics[scale=0.6]{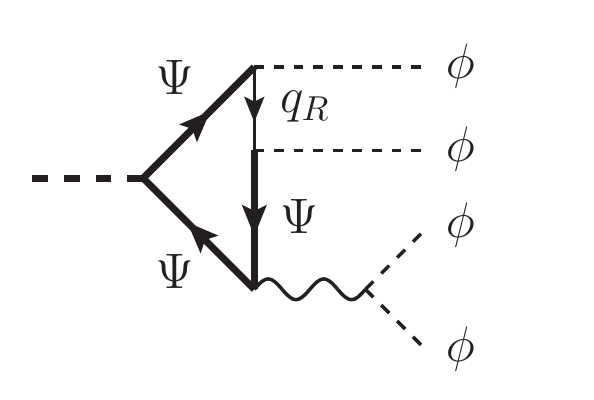}
\includegraphics[scale=0.6]{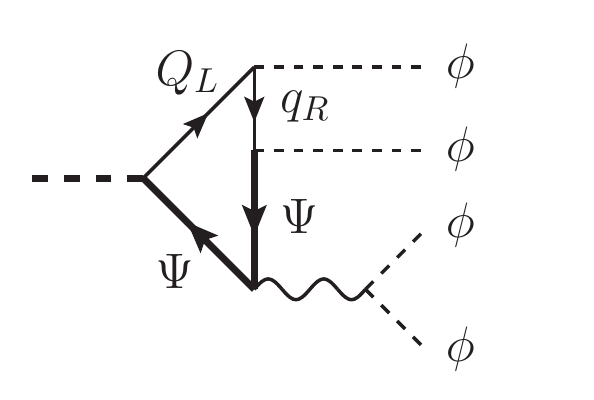}
\includegraphics[scale=0.6]{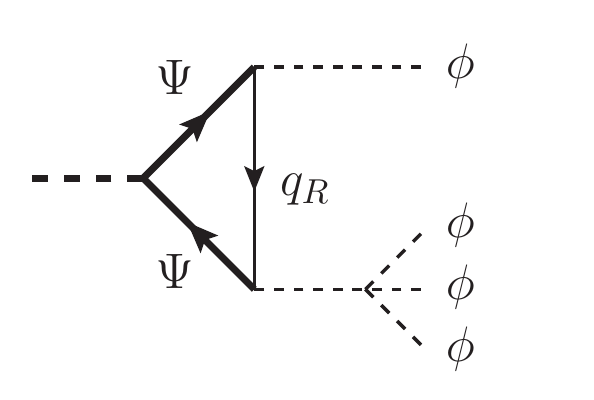}
\includegraphics[scale=0.6]{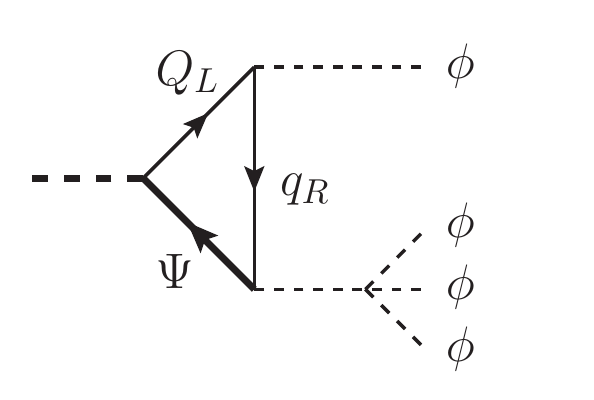}
\end{center}
\vspace{-5mm}
\caption{\label{fig:diagrams2} 
Example diagrams contributing to the matching of the Wilson coefficient $\widetilde C_{\phi\phi\phi\phi}$.}
\end{figure*}

The relevant ${\mathcal O}(\lambda^4)$ operators in the \scetbsm{} Lagrangian mediating $S\to Zh$ decay can be written in the form \cite{Alte:2018nbn}
\begin{equation}\label{Leff4}
   {\mathcal L}_{\rm eff}^{(4)} 
   = \frac{\widetilde C_{\phi\phi\phi\phi}(M,M_S)}{M}\,2iS_v\!
    \left( \Phi_{n_1}^\dagger \Phi_0\,\Phi_{n_2}^\dagger \Phi_0 - \mathrm{h.c.} \right) + \dots \,,
\end{equation}
where the dots stand for fermionic operators, which contribute to the decay amplitude at one-loop order. The operator written out explicitly gives the tree-level contribution
\begin{equation}\label{SZhampl}
   {\mathcal M}(S\to Z_\parallel h) \big|_{\rm tree} 
   = - i\hspace{0.3mm}\widetilde C_{\phi\phi\phi\phi}\,\frac{v^2}{M} \,.
\end{equation}
Since in the VLQ model we consider the Wilson coefficient $\widetilde C_{\phi\phi\phi\phi}$ is generated starting at one-loop order, it will be necessary to include other loop-suppressed effects for consistency (see below). 

Representative one-loop diagrams contributing to the matching coefficient $\widetilde C_{\phi\phi\phi\phi}$ are shown in Figure~\ref{fig:diagrams2}. Evaluating the relevant graphs in the $\overline{\rm MS}$ scheme, we obtain after a lengthy calculation
\begin{widetext}
\begin{equation}\label{Cphi4res}
\begin{aligned}
   \widetilde C_{\phi\phi\phi\phi} 
   &= - \frac{N_c}{16\pi^2\xi} \sum_{q=u,d} 2T_3^q\,\Bigg\{ 
    \mbox{Tr}\big(\tilde{\bm{X}}\bm{G}_q\bm{Y}_q^\dagger\bm{Y}_q\bm{G}_q^\dagger\big)\,
    \Big[ - L_M \Big( \xi + (1-\xi) \ln(1-\xi) \Big) + f_1(\xi) \Big] \\
   &\qquad\mbox{}+ \mbox{Tr}\big(\tilde{\bm{X}}\bm{G}_q\bm{G}_q^\dagger\bm{G}_q\bm{G}_q^\dagger\big)\,f_2(\xi) 
    + \mbox{Tr}\big(\tilde{\bm{X}}\bm{G}_q\bm{G}_q^\dagger\big)\,
    \bigg[ (g^2+g^{\prime\,2})\,f_3(\xi) + \lambda_H\,f_4(\xi) \bigg] \\
   &\qquad\mbox{}+ \mbox{Im\,Tr}\big(\bm{G}_q^\dagger\bm{V}_Q\bm{Y}_q\bm{Y}_q^\dagger\bm{Y}_q\big)\,    
    \bigg[ \frac{\xi}{2}\,L_S^2 - L_M \Big( \xi + (1+\xi) \ln(1-\xi) \Big) + f_5(\xi) \bigg] \\
   &\qquad\mbox{}+ \mbox{Im\,Tr}\big(\bm{G}_q^\dagger\bm{V}_Q\bm{Y}_q\bm{G}_q^\dagger\bm{G}_q\big)\,
    \Big[ \xi \left( L_S - L_M \right) \ln(1+\xi) + f_6(\xi) \Big] \\
   &\qquad\mbox{}+ \mbox{Im\,Tr}\big(\bm{G}_q^\dagger\bm{V}_Q\bm{Y}_q\big)\,\bigg[ 
    (g^2+g^{\prime\,2})\,\Big[ - L_M \Big(1 + \frac{4-3\xi}{4\xi} \ln(1-\xi) \Big)
    - \frac{\xi}{4} \left( L_S - L_M \right) + f_7(\xi) \Big] \\
   &\hspace{38mm}\mbox{}+ \lambda_H\,\Big[ 2 L_M \ln(1-\xi) - 2\xi \left( L_S - L_M \right) + f_8(\xi) \Big] 
    \bigg] \Bigg\} \\
   &\quad\mbox{}- \frac{N_c}{16\pi^2\xi} \sum_{q=u,d} Q_q\,g^{\prime\,2}\,\Bigg\{
    \mbox{Tr}\big(\tilde{\bm{X}}\bm{G}_q\bm{G}_q^\dagger\big)\,
    \bigg[ L_M \left( \frac{2-\xi}{2} + \frac{1-\xi}{\xi} \ln(1-\xi) \right) + f_9(\xi) \bigg] \\
   &\qquad\mbox{}+ \mbox{Im\,Tr}\big(\bm{G}_q^\dagger\bm{V}_Q\bm{Y}_q\big)\,
    \bigg[ L_M \left( \frac{6-\xi}{2} + \frac{3-2\xi}{\xi} \ln(1-\xi) \right) + f_{10}(\xi) \bigg] \Bigg\} \,.
\end{aligned}
\end{equation}
\end{widetext}
Here $T_3^u=\frac12$ and $T_3^d=-\frac12$ are the weak isospins of up- and down-type quarks, $Q_q$ denote the quark electric charges in units of $e$, $\lambda_H$ is the quartic coupling of the Higgs field, and $g$, $g'$ are the gauge couplings of $SU(2)_L$ and $U(1)_Y$. The logarithms $L_M=\ln(M^2/\mu^2)$ and $L_S=\ln(M_S^2/\mu^2)-i\pi$ contain the dependence on the factorization scale $\mu$, and we have defined the functions $f_i(\xi)$ collected in Appendix~\ref{app:B}. For $\xi>1$, the above expressions must be analytically continued using the prescription $\xi\to\xi+i\epsilon$. In the limit where $\xi\ll 1$, corresponding to $M_S^2\ll M^2$, the result (\ref{Cphi4res}) can be expanded in powers of $\xi$. We find that the leading terms of ${\mathcal O}(\xi^0)$ agree with eq.\,(6.20) in \cite{Alte:2018nbn}, where we had defined the matrices $\hat{\bm{Y}}_q=\bm{G}_q^\dagger\bm{V}_Q$. Moreover, the terms linear in $\xi$ are consistent with eq.\,(24) in \cite{Bauer:2016ydr}.

An interesting feature of the result (\ref{Cphi4res}) is the rather complicated dependence of the terms involving the factorization scale $\mu$, which are contained in the logarithms $L_M$ and $L_S$, on the mass ratio $\xi=M_S^2/M^2$. In conventional EFT applications the coefficients of the $\mu$-dependent terms in Wilson coefficients and operator matrix elements are functions of the coupling constants of the theory, but they do not depend in a non-trivial way on the masses of heavy particles that have been integrated out. The reason is that the $\mu$-dependence must cancel between Wilson coefficients and matrix elements, and the low-energy theory does not know about the masses of the heavy particles. 

In the present case, the $\mu$-dependence of the contribution to the $S\to Zh$ decay amplitude entering via the Wilson coefficient $\widetilde C_{\phi\phi\phi\phi}$ in (\ref{Cphi4res}) is cancelled by the scale dependence of one-loop matrix elements of operators involving fermion pairs, which are induced by tree-level matching at ${\mathcal O}(\lambda^4)$. Indeed, since in our model $\widetilde C_{\phi\phi\phi\phi}$ arises at one-loop order, the one-loop matrix elements of other ${\mathcal O}(\lambda^4)$ operators, which appear already at tree level, must be included for consistency. The relevant terms can be extracted from (\ref{LlinearS}). For the purpose of illustration we consider the last operator in this result, which contains the flavor matrices $\bm{X}$ and $\tilde{\bm{X}}$. At ${\mathcal O}(\lambda^4)$, it gives rise to the structure
\begin{equation}
\begin{aligned}
   {\mathcal L}_\mathrm{eff} \big|_{S^1}^{\lambda^4}
   &\ni - \frac{1}{M} \sum_{q=u,d} \bigg[ \bar q_{R,n_1} \Phi_0^\dagger\,S_v\,
    \bm{G}_q^\dagger \big( \bm{X} - i\gamma_5\tilde{\bm{X}} \big) \bm{G}_q \\
   &\quad\times \frac{1}{M^2+\Box}\,i\delsl\,\Phi_{n_2}\,q_{R,n_1} + \mathrm{h.c.} \bigg] + (n_1\leftrightarrow n_2) \,.
\end{aligned}
\end{equation}
For $q=u$ the doublet $\Phi$ must be replaced by $\tilde\Phi$. We only need to consider operators where both fermions are described by collinear fields moving along the same direction, since later we need to take matrix elements where the fermion pair is converted into a collinear Higgs or $Z$ boson. Between the collinear spinors only the $n_1\cdot\partial$ component of the derivative survives, and hence the derivative gives zero when acting on the fermions. We now define the following set of \scetbsm{} hermitian operators (here and below we abbreviate $\bar u\equiv 1-u$):
\begin{equation}\label{opsdef}
\begin{aligned}
   O_{q_R\bar q_R\phi\phi}^{(\pm)\,ij}(u)
   &= S_v \left[ \bar q_{R,n_1}^{(u)\,i}\,\vsl\,q_{R,n_1}^{(\bar u)\,j} 
    \mp \bar q_{R,n_1}^{(\bar u)\,i}\,\vsl\,q_{R,n_1}^{(u)\,j} \right] \\
   &\quad\times \left( \Phi_{n_2}^\dagger \Phi_0 \pm \Phi_0^\dagger \Phi_{n_2} \right) , \\
   \widetilde O_{q_R\bar q_R\phi\phi}^{(\pm)\,ij}(u)
   &= i S_v \left[ \bar q_{R,n_1}^{(u)\,i}\,\vsl\,q_{R,n_1}^{(\bar u)\,j} 
    \pm \bar q_{R,n_1}^{(\bar u)\,i}\,\vsl\,q_{R,n_1}^{(u)\,j} \right] \\
   &\quad\times \left( \Phi_{n_2}^\dagger \Phi_0 \pm \Phi_0^\dagger \Phi_{n_2} \right) .
\end{aligned}
\end{equation}
Here $u$ denotes the fraction of the total $n_1$-collinear momentum carried by the final-state quark. The operators shown in the first line are CP even, while those in the second line are CP odd. Writing the relevant terms in the Lagrangian in the form
\begin{equation}
\begin{aligned}
   {\mathcal L}_{\rm eff}^{(4)}
   &\ni \frac{1}{M^2} \sum_{q=u,d} \int_0^1\!du\,\bigg[ 
    C_{q_R\bar q_R\phi\phi}^{(\pm)\,ij}(M,M_S,u)\,O_{q_R\bar q_R\phi\phi}^{(\pm)\,ij}(u) \\
   &\quad\mbox{}+ \widetilde C_{q_R\bar q_R\phi\phi}^{(\pm)\,ij}(M,M_S,u)\,
    \widetilde O_{q_R\bar q_R\phi\phi}^{(\pm)\,ij}(u) \bigg] \,,
\end{aligned}
\end{equation}
we obtain the Wilson coefficients (in matrix notation)
\begin{equation}\label{eq29}
\begin{aligned}
   \bm{C}_{q_R\bar q_R\phi\phi}^{(+)}(u,M,M_S)
   &= \frac{M_S}{2M}\,\frac{\bm{G}_q^\dagger \bm{X} \bm{G}_q}{1-\xi\bar u} \,, \\
   \bm{C}_{q_R\bar q_R\phi\phi}^{(-)}(u,M,M_S)
   &= \frac{M_S}{2M}\,2T_3^q\,\frac{\bm{G}_q^\dagger \bm{X} \bm{G}_q}{1-\xi\bar u} \,, \\
   \widetilde{\bm{C}}_{q_R\bar q_R\phi\phi}^{(+)}(u,M,M_S)
   &= \frac{M_S}{2M}\,\frac{\bm{G}_q^\dagger \tilde{\bm{X}} \bm{G}_q}{1-\xi\bar u} \,, \\
   \widetilde{\bm{C}}_{q_R\bar q_R\phi\phi}^{(-)}(u,M,M_S)
   &= \frac{M_S}{2M}\,2T_3^q\,\frac{\bm{G}_q^\dagger \tilde{\bm{X}} \bm{G}_q}{1-\xi\bar u} \,.
\end{aligned}
\end{equation}
The factors $2T_3^q$ arise because for $q=u$ the operators involve the scalar doublets $\tilde\Phi$ rather than $\Phi$.

\begin{figure}
\begin{center}
\includegraphics[scale=0.57]{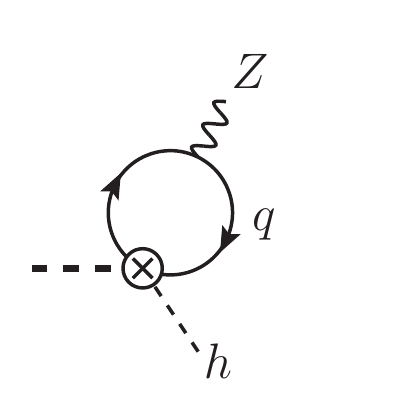}
\includegraphics[scale=0.57]{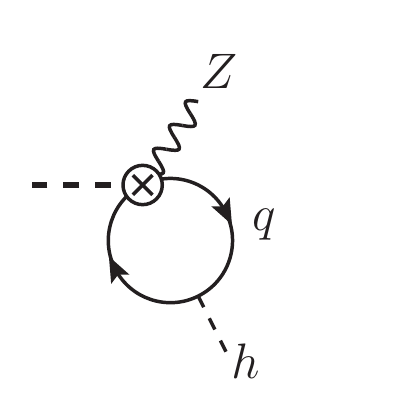}
\end{center}
\vspace{-5mm}
\caption{\label{fig:lam4ops} 
One-loop contributions of the operators $\widetilde O_{q_R\bar q_R\phi\phi}^{(+)\,ij}$ (left) and $\widetilde O_{q_R\bar q_R\phi\phi}^{(-)\,ij}$ (right) to the $S\to Zh$ decay amplitude.}
\end{figure}

The CP-odd operators in (\ref{opsdef}) contribute at one-loop order to the $S\to Zh$ decay amplitude via the diagrams shown in Figure~\ref{fig:lam4ops}. Working in the fermion mass basis, we find in the $\overline{\rm MS}$ scheme 
\begin{equation}\label{eq31}
\begin{aligned}
   &i\hspace{0.3mm}\langle Z_\parallel h|\,\widetilde O_{q_R\bar q_R\phi\phi}^{(+)\,ij}(u)\,|S\rangle \\
   &= \frac{\delta^{ij} N_c}{8\pi^2}\,v^2 M_S \big( y_q^2\,T_3^q - Q_q g^{\prime\,2} u\bar u \big)
    \ln\frac{\mu^2}{m_q^2-u\bar u m_Z^2-i\epsilon} \,, \\
   &i\hspace{0.3mm}\langle Z_\parallel h|\,\widetilde O_{q_R\bar q_R\phi\phi}^{(-)\,ij}(u)\,|S\rangle \\
   &= \frac{\delta^{ij} N_c}{16\pi^2}\,v^2 M_S\,y_q^2\,(u-\bar u)
    \ln\frac{\mu^2}{m_q^2-u\bar u m_h^2-i\epsilon} \,. 
\end{aligned}
\end{equation}
Multiplying these expressions with the corresponding Wilson coefficients in (\ref{eq29}) and integrating the result over $u$ we obtain the contribution to the $S\to Zh$ decay amplitude, which must be added to the one in (\ref{SZhampl}).

Here we are mainly concerned with the cancellation of the $\mu$-dependent terms in the final expression for the decay amplitude. Note that the scale-dependent terms in (\ref{eq31}) have simple coefficients involving coupling constants and some factors of $v$ and $M_S$ needed for dimensional reasons. The non-trivial dependence on the mass ratio $\xi$ arises when these matrix elements are multiplied by the corresponding Wilson coefficients and integrated over the variable $u$. To display our results we use the $Z$-boson mass in the denominator of the corresponding logarithms, and we omit the remaining terms that are scale independent and free of large logarithms. Combining the contributions in (\ref{SZhampl}) and (\ref{eq31}), we find
\begin{widetext}
\begin{equation}
\begin{aligned}
   \langle Z_\parallel h|\,{\mathcal L}_{\rm eff}^{(4)}\,|S\rangle 
   &= - i\hspace{0.3mm}\widetilde C_{\phi\phi\phi\phi}\,\frac{v^2}{M} 
    - i\hspace{0.3mm}\frac{N_c}{16\pi^2\xi}\,\frac{v^2}{M} \sum_{q=u,d} \bigg\{
    2T_3^q\,\mbox{Tr}\big(\tilde{\bm{X}}\bm{G}_q\bm{Y}_q^\dagger\bm{Y}_q\bm{G}_q^\dagger\big) 
    \ln\frac{\mu^2}{m_Z^2} \Big( \xi + (1-\xi) \ln(1-\xi) \Big) \\
   &\hspace{55mm}\mbox{}- Q_q\,g^{\prime\,2}\,\mbox{Tr}\big(\tilde{\bm{X}}\bm{G}_q\bm{G}_q^\dagger\big) 
    \ln\frac{\mu^2}{m_Z^2} \bigg( \frac{2-\xi}{2} + \frac{1-\xi}{\xi} \ln(1-\xi) \bigg) \\
   &\hspace{55mm}\mbox{}+ \mbox{terms involving $\bm{V}_Q$} \bigg\} 
    + \mbox{scale-independent terms} \,.
\end{aligned}
\end{equation}
We have transformed the expressions (\ref{eq31}) back to the weak basis by replacing $y_q^2\,\delta_{ij}\to(\bm{Y}_q^\dagger\bm{Y}_q)_{ij}$. Inspection of (\ref{Cphi4res}) shows that the $\mu$-dependent terms indeed cancel out in this result.
\end{widetext}

\section{Resummation of large logarithms}
\label{sec:pheno}

\scetbsm{} offers a systematic framework for expanding the decay amplitudes for the heavy resonance $S$ into SM particles in powers of $v/M_S$ and resumming large logarithms of this scale ratio. (As before, we assume that the scales $M$ and $M_S$ are of similar magnitude.) Since the rates are affected by Sudakov double logarithms, resummation is important even in cases where the logarithms arise from electroweak interactions \cite{Chiu:2007yn,Chiu:2007dg,Chiu:2008vv}. These logarithms suppress the decay rates and hence should be taken into account when deriving bounds on the masses and couplings of hypothetical new heavy particles. We now illustrate this point by focussing on a few important two-body decay modes of a heavy scalar resonance $S$. 

For the purposes of illustration, we assume $M_S=2$\,TeV and $M=2.5$\,TeV for the masses of $S$ and of the VLQs, respectively. We calculate the Wilson coefficients in the effective Lagrangians (\ref{Leff2}) and (\ref{Leff3}) at the high scale $\mu=M$ and evolve them down to a characteristic scale for the process of interest. This evolution is governed by renormalization-group (RG) equations derived in \cite{Alte:2018nbn}. As long as the characteristic scale is of the order of the weak scale, it is appropriate to include all SM particles in the anomalous dimensions and $\beta$-functions of the EFT. A consistent approximation is obtained by including the leading terms in the matching coefficients at the high scale and using two-loop approximations for the cusp anomalous dimension and $\beta$-functions as well as one-loop approximations for all other anomalous dimensions in the evolution to low energies (see below).

\subsection{\boldmath $S\to\mathrm{2\,jets}$ decay}

At lowest order in perturbation theory the process $S\to\mathrm{2\,jets}$ proceeds primarily via the decay $S\to gg$, whose rate is enhanced by a factor $M_S^2/v^2$ relative to the $S\to q\bar q$ decay rate. Also, in many models the latter rate is suppressed by the light quark masses. We thus obtain $\Gamma(S\to\mbox{2\,jets})\approx\Gamma(S\to gg)$ with
\begin{equation}\label{Gamma2jets}
   \Gamma(S\to gg) 
   = \frac{M^2}{M_S}\,8\pi\alpha_s^2(\mu_j) \left( |C_{GG}(\mu_j)|^2 + |\widetilde C_{GG}(\mu_j)|^2 \right) .
\end{equation}
Here $\mu_j$ is the characteristic scale inherent in the definition of the jets, such as an upper bound on the jet invariant mass. At the high matching scale $\mu_h=M$ the relevant Wilson coefficients have been given in (\ref{eq:cAA}). The two coefficients obey the same RG equation \cite{Alte:2018nbn}
\begin{equation}
   \mu\,\frac{d}{d\mu}\,C_{GG}(\mu) 
   = \left[ 3\gamma_{\rm cusp}^{(3)} \left( \ln\frac{M_S^2}{\mu^2} - i\pi \right) + 2\gamma^G \right] 
    C_{GG}(\mu) \,.
\end{equation}
Note the important fact that for Sudakov problems the anomalous dimensions themselves contain a (so-called ``cusp'') logarithm, and that they have non-zero imaginary parts. At leading logarithmic order, we need $\gamma_{\rm cusp}^{(3)}$ to two-loop and $\gamma^G$ to one-loop order. The relevant expressions are
\begin{equation}
   \gamma_{\rm cusp}^{(3)} 
   = \frac{\alpha_s}{\pi} + \left( \frac{47}{12} - \frac{\pi^2}{4} \right) \left( \frac{\alpha_s}{\pi} \right)^2 
    + {\mathcal O}(\alpha_s^3) \,, 
\end{equation}
and $\gamma^G=0+{\mathcal O}(\alpha_s^2)$. Solving the RG equation, we find $C_{GG}(\mu)/C_{GG}(M)=\widetilde C_{GG}(\mu)/\widetilde C_{GG}(M)=U_{GG}(\mu,M)$, where \cite{Becher:2006nr,Becher:2007ty} 
\begin{equation}\label{UGGres}
   U_{GG}(\mu,M)
   = \exp\left[ \frac{6}{49}\,g(\mu,M) + \frac67 \left( i\pi - \ln\xi \right) \ln r \right] ,
\end{equation}
with $r=\alpha_s(\mu)/\alpha_s(M)$ and 
\begin{equation}
\begin{aligned}
   g(\mu,M) &= - \frac{4\pi}{\alpha_s(M)} \left( \frac{1}{r} -1 + \ln r \right) \\
   &\quad\mbox{}- \left( \frac{251}{21} - \pi^2 \right) \big( r -1 - \ln r \big)
    + \frac{13}{7}\,\ln^2 r \,.
\end{aligned}
\end{equation}
An analogous relation holds for $\widetilde C_{GG}$. If we assume that the characteristic jet scale is $\mu_j=100$\,GeV, then
\begin{equation}
   U_{GG}(\mu_j,M)
   \approx 0.38\,e^{0.98\hspace{0.2mm}i} \,.
\end{equation}
The decay rate in (\ref{Gamma2jets}) is suppressed by the factor $|U_{GG}(\mu_j,M)|^2\approx 0.147$. Not including these resummation effects would vastly overestimate the decay rate.

\subsection{\boldmath $S\to t\bar t$ decay}

The largest two-body decay rate into quark-antiquark final states is likely to be that into top quarks. At leading order in perturbation theory the corresponding decay rate is given by 
\begin{equation}\label{Gammatt}
   \Gamma(S\to t\bar t) 
   = \frac{3}{16\pi}\,\frac{v^2 M_S}{M^2}\,\sqrt{1-\frac{4m_t^2}{M_S^2}}\,
   \left| \big({\bf C}_{Q_L \bar u_R}\big)_{33}(m_t) \right|^2 .
\end{equation}
At the high matching scale $\mu_h=M$ the coefficient $\bm{C}_{Q_L \bar u_R}$ has been given in (\ref{eq:twoJetLambda3}). The related coefficient ${\bf C}_{Q_L \bar u_R}$ (with a straight letter ``C'') is obtained by transforming this expression to the quark mass basis. Including only QCD effects, it obeys the RG equation \cite{Alte:2018nbn}
\begin{equation}
\begin{aligned}
   & \mu\,\frac{d}{d\mu}\,{\bf C}_{Q_L \bar u_R}(\mu) \\
   &= \left[ \frac43\gamma_{\rm cusp}^{(3)} \left( \ln\frac{M_S^2}{\mu^2} - i\pi \right) + 2\gamma^q \right] 
    {\bf C}_{Q_L \bar u_R}(\mu) \,,
\end{aligned}
\end{equation}
where $\gamma^q=-\alpha_s/\pi+{\mathcal O}(\alpha_s^2)$. Solving this equation, we obtain ${\bf C}_{Q_L \bar u_R}(\mu)=U_{q\bar q}(\mu,M)\,{\bf C}_{Q_L \bar u_R}(M)$ with
\begin{equation}
   U_{q\bar q}(\mu,M)
   = \exp\!\left[ \frac{8}{147}\,g(\mu,M) + \frac{8}{21}\!\left(\! i\pi + \frac32 - \ln\xi \right)\!\ln r \right]\! .
\end{equation}
Evolving the coefficient down to the scale of the top-quark mass, we find
\begin{equation}
   U_{q\bar q}(m_t,M)
   \approx 0.90\,e^{0.31\hspace{0.2mm}i} \,.
\end{equation}
The decay rate in (\ref{Gammatt}) is suppressed by the factor $|U_{q\bar q}(m_t,M)|^2\approx 0.81$. In this case, resummation effects have a more modest impact on the decay rate.

\subsection{\boldmath $S\to\gamma\gamma$ decay}

It is instructive to also consider an example where only electroweak Sudakov logarithms contribute. The diphoton decay mode has a very similar structure as the $S\to gg$ mode discussed above. At leading order in perturbation theory the decay rate is given by 
\begin{equation}\label{Gamma2gamma}
\begin{aligned}
   \Gamma(S\to\gamma\gamma) 
   = \frac{M^2}{M_S}\,\pi\alpha^2 \Big( & |C_{WW}(m_W) + C_{BB}(m_W)|^2 \\
   + & |\widetilde C_{WW}(m_W) + \widetilde C_{BB}(m_W)|^2 \Big) \,.
\end{aligned}
\end{equation}
Here $\alpha\approx 1/137.036$ is the fine-structure constant. The Wilson coefficients need to be evolved down to the scale of electroweak symmetry breaking, which we identify with the mass of the $W$ boson. Below the weak scale the running stops. At the high matching scale $\mu_h=M$ the relevant coefficients have been given in (\ref{eq:cAA}). The coefficients $C_{WW}$ and $\widetilde C_{WW}$ obey the same RG equation \cite{Alte:2018nbn}
\begin{equation}
   \mu\,\frac{d}{d\mu}\,C_{WW}(\mu) 
   \!=\! \left[ 2\gamma_{\rm cusp}^{(2)}\!\left(\! \ln\frac{M_S^2}{\mu^2} - i\pi \right) + 2\gamma^W \!\right]\!
    C_{WW}(\mu) \,.
\end{equation}
The relevant cusp anomalous dimension is
\begin{equation}\label{gam2}
   \gamma_{\rm cusp}^{(2)} 
   = \frac{\alpha_2}{\pi} + \left( 2 - \frac{\pi^2}{6} \right) \left( \frac{\alpha_2}{\pi} \right)^2 + \dots \,, 
\end{equation}
whereas $\gamma^W$ vanishes at one-loop order. Here $\alpha_2=g^2/(4\pi)$ is the coupling constant of $SU(2)_L$. The Wilson coefficients $C_{BB}$ and $\widetilde C_{BB}$, on the other hand, are scale independent at leading logarithmic order. It follows that
\begin{equation}
\begin{aligned}
   & C_{WW}(m_W) + C_{BB}(m_W) \\
   &= U_{WW}(m_W,M)\,C_{WW}(M) + U_{BB}(m_W,M)\,C_{BB}(M)
\end{aligned}
\end{equation}
and similarly for the other two coefficients in (\ref{Gamma2gamma}), where $U_{BB}(m_W,M)\approx 1$, while $U_{WW}(\mu,M)$ is given by an expression similar to (\ref{UGGres}), but with different numerical coefficients and with $\alpha_s(\mu)$ replaced by the coupling $\alpha_2(\mu)$. Numerically, we obtain
\begin{equation}
   U_{WW}(m_W,M) \approx 0.80\,e^{0.23\hspace{0.2mm}i} \,, \quad
   U_{BB}(m_W,M) \approx 1 \,.
\end{equation}
The impact of these resummation effects on the diphoton decay rate depends on the values of $\kappa_1/M$ and $\mathrm{Tr}(\bm{X})$ in (\ref{eq:cAA}). In the limit where the term proportional to $\kappa_1$ can be neglected, the decay rate is suppressed by the factor $|0.9\,U_{WW}(m_W,M)+0.1|^2\approx 0.67$. The resummation of electroweak Sudakov logarithms thus has a sizable impact on the rate.

\subsection{\boldmath $S\to hh$ decay}

As a final example we consider the decay mode $S\to hh$, whose rate is given by 
\begin{equation}
   \Gamma_{S\to hh} = \frac{M^2}{32\pi M_S}\,\sqrt{1-\frac{4m_h^2}{M_S^2}}\,|C_{\phi\phi}(m_h)|^2 \,.
\end{equation}
The Wilson coefficient satisfies the RG equation \cite{Alte:2018nbn}
\begin{equation}\label{eq49}
\begin{aligned}
   & \mu\,\frac{d}{d\mu}\,C_{\phi\phi}(\mu) \\
   &= \left[ \left( \frac14\,\gamma_{\rm cusp}^{(1)} + \frac34\,\gamma_{\rm cusp}^{(2)} \!\right)
    \left( \ln\frac{M_S^2}{\mu^2} - i\pi \right) + 2\gamma^\phi \right] C_{\phi\phi}(\mu) \,,
\end{aligned}
\end{equation}
where
\begin{equation}
\begin{aligned}
   \gamma_{\rm cusp}^{(1)} &= \frac{\alpha_1}{\pi} - \frac{17}{6} \left( \frac{\alpha_1}{\pi} \right)^2 + \dots \,, \\
   \gamma^\phi &= - \frac{\alpha_1}{4\pi} - \frac{3\alpha_2}{4\pi} + \frac{3y_t^2}{8\pi^2} + \dots \,,
\end{aligned}
\end{equation}
and $\gamma_{\rm cusp}^{(2)}$ has been given in (\ref{gam2}). Here $\alpha_1$ is the coupling constant of $U(1)_Y$ (not rescaled by a factor 5/3). Since there are now three different couplings involved, it is easiest to integrate the RG equation (\ref{eq49}) numerically, using the one-loop $\beta$-functions for the various couplings. Writing the solution in the form $C_{\phi\phi}(m_h)=U_{\phi\phi}(m_h,M)\,C_{\phi\phi}(M)$, we find 
\begin{equation}
   U_{\phi\phi}(m_h,M) \approx 0.79\,e^{0.08\hspace{0.2mm}i}  \,.
\end{equation}
It follows that the di-Higgs decay rate is suppressed by the factor $|U_{\phi\phi}(m_h,M)|^2\approx 0.62$, which is once again a significant correction.

\section{Conclusions}

When a new heavy resonance beyond the SM is discovered, it will be important to have an effective field-theory description of its decay and production modes, in which the new-physics scale $M$ is disentangled from the electroweak scale. This effective theory should be able to deal with the situation that the new state is a member of a larger sector of new physics. In this paper we have illustrated the recently developed \scetbsm{} approach \cite{Alte:2018nbn} to solve this problem in the context of an extension of the SM by a heavy scalar singlet $S$ and a set of vector-like heavy quarks. We have performed the matching calculation for the Wilson coefficients in the effective Lagrangian both at tree level and including the leading one-loop corrections. These coefficients are in general non-trivial functions of the mass ratio $\xi=M_S^2/M^2$, where $M_S$ is the mass of the scalar resonance while $M$ sets the masses of the vector-like quarks. In this way, our effective theory resums an infinite tower of local operators in the conventional effective field-theory approach to describe the interactions of $S$ with SM fields. For the special case of the decay $S\to Zh$, the Wilson coefficient of the relevant operator contains logarithms of the form $\ln(M_S^2/\mu^2)$ and $\ln(M^2/\mu^2)$ with coefficients that depend in a non-polynomial way on the ratio $\xi$. We have explained the origin of this effect and demonstrated how the scale dependence is cancelled in the effective theory.

The \scetbsm{} framework allows one to resum large Sudakov logarithms affecting the decay rates of $S$ into SM particles. We have explicitly performed the resummation at leading logarithmic order for the decays $S\to\mathrm{2\,jets}$, $S\to t\bar t$, $S\to\gamma\gamma$ and $S\to hh$, finding that in all cases the decay rates are significantly reduced. It is important to take these resummation effects into account when placing bounds on the masses and couplings of hypothetical new heavy particles. Possible avenues worthy to pursue in the future include extensions of our work to resonances of non-zero spin as well as particles that are not singlets under the SM gauge group. In this way, the \scetbsm{} approach can be applied to collider searches for heavy particles proposed in many extensions of the SM.

\begin{acknowledgments}
We are grateful to Florian Goertz for useful discussions. The work of S.A.\ and M.N.\ was supported by the DFG Clusters of Excellence {\em Precision Physics, Fundamental Interactions and Structure of Matter\/} PRISMA (EXC~1098) and PRISMA$^+$ (EXC~2118). The research of S.A.\ was also supported by the DFG Research Training Group {\em Symmetry Breaking in Fundamental Interactions\/} (GRK~1581). M.K.\ gratefully acknowledges support by the Swiss National Science Foundation (SNF) under contract 200021-175940.
\end{acknowledgments}

\appendix

\section{\boldmath Calculation of the quantity $\delta_{\kappa_1}$}
\label{app:A}

\begin{figure}
\begin{center}
\vspace{10mm}
\raisebox{3mm}{\includegraphics[scale=0.44]{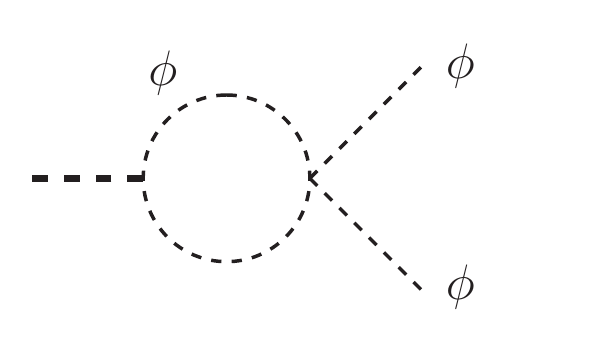}}
\hspace{-5mm}
\includegraphics[scale=0.44]{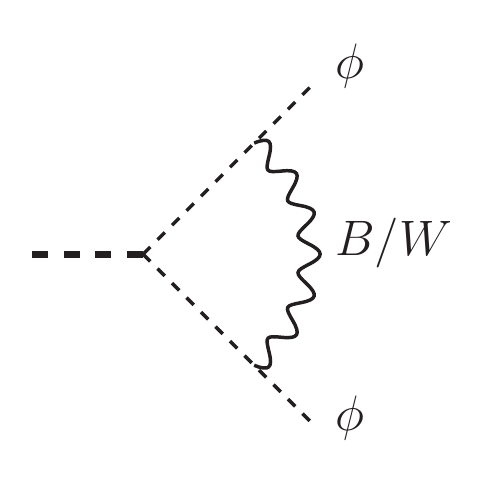}
\includegraphics[scale=0.44]{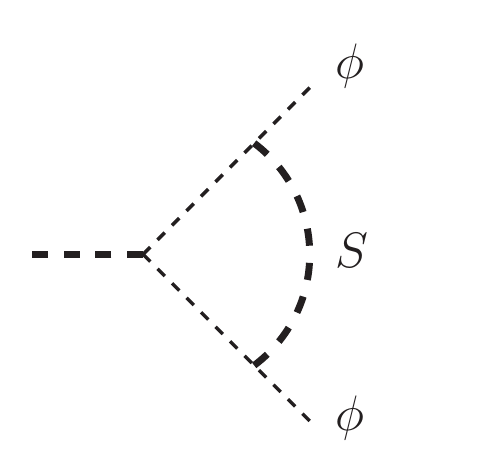}
\hspace{-5mm}
\includegraphics[scale=0.44]{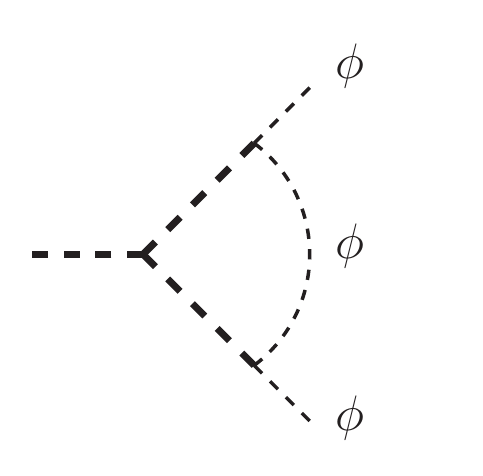}
\includegraphics[scale=0.46]{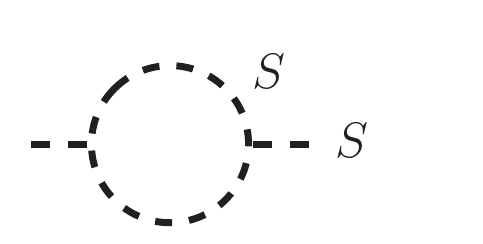}
\hspace{-3.5mm}
\includegraphics[scale=0.46]{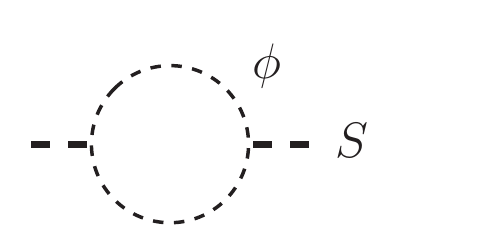}
\hspace{-3.5mm}
\includegraphics[scale=0.46]{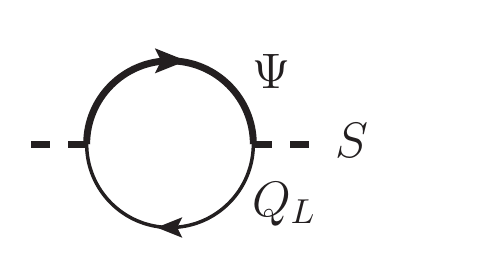}
\hspace{-3.5mm}
\includegraphics[scale=0.46]{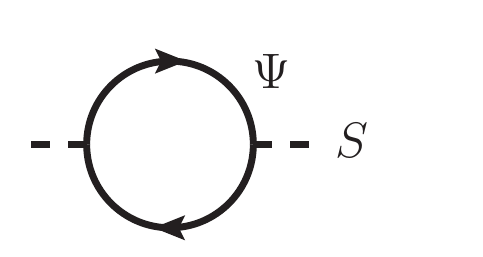}
\raisebox{-10mm}{\includegraphics[scale=0.45]{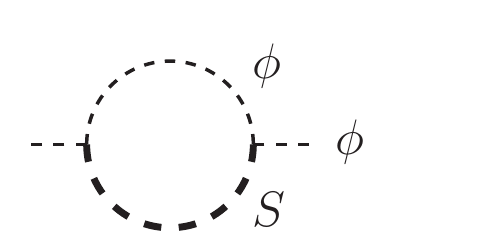}
\hspace{-3mm}
\includegraphics[scale=0.45]{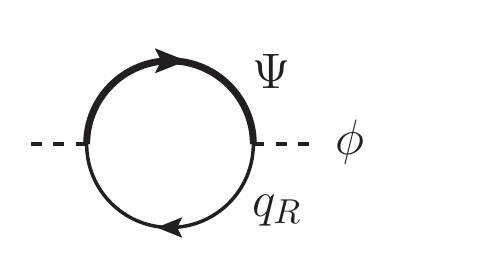}}
\end{center}
\caption{\label{fig:evenmorediagrams} 
One-loop diagrams contributing to the parameter $\delta_{\kappa_1}$. The graphs in the last two lines show the matching corrections to the wave-function renormalization constants of the heavy resonance $S$ and the Higgs scalar $\phi$.}
\end{figure}

Here we report our result for the one-loop coefficient $\delta_{\kappa_1}$ in (\ref{eq18}). It receives contributions from the vertex-correction diagrams shown in the first line of Figure~\ref{fig:evenmorediagrams} as well as from hard matching corrections to the wave-function renormalization constants of the scalar fields. We find
\begin{widetext}
\begin{equation}
\begin{aligned}
   \delta_{\kappa_1} 
   &= \frac{3\lambda_H}{8\pi^2} \left( L_S - 2 \right) 
   - \frac{3g^2+g^{\prime 2}}{64\pi^2} \left( L_S^2 - L_S + 2 - \frac{\pi^2}{6} \right) 
    + \frac{\kappa_1^2}{16\pi^2 M_S^2} \left( \frac12 + \frac{\pi^2}{12} + i\pi\ln 2 \right) 
    + \frac{\kappa_1\lambda_3}{16\pi^2 M_S^2}\,\frac{\pi^2}{9} \\
   &\quad\mbox{}- \frac{\lambda_3^2}{64\pi^2 M_S^2} \left( \frac{2\pi}{3\sqrt3} - 1 \right) 
    + \frac{N_c}{8\pi^2}\,\mbox{Tr}\big(\bm{V}_Q^\dagger\bm{V}_Q\big)
    \left[ L_M - 1 - \frac{1}{\xi} - \frac{1-\xi^2}{\xi^2} \ln(1-\xi) \right] \\
   &\quad\mbox{}+ \frac{N_c}{8\pi^2}\,\mbox{Tr}\big(\bm{X}^2\big) \left[ L_M - 1 - \frac{4}{\xi}
    + \frac{2(2+\xi)}{\xi} \sqrt{\frac{4-\xi}{\xi}}\,g(\xi) \right]
    + \frac{N_c}{8\pi^2}\,\mbox{Tr}\big(\tilde{\bm{X}}^2\big) \left[ L_M - 1 
    + \frac{2(2-\xi)}{\sqrt{\xi(4-\xi)}}\,g(\xi) \right] \\
   &\quad\mbox{}+ \frac{N_c}{8\pi^2}\,\mbox{Tr}\big(\bm{G}_u^\dagger\bm{G}_u+\bm{G}_d^\dagger\bm{G}_d\big)
    \left( L_M - \frac12 \right) ,
\end{aligned}
\end{equation}
where $L_M=\ln(M^2/\mu^2)$ and $L_S=\ln(M_S^2/\mu^2)-i\pi$, and the function $g(\xi)$ has been given in (\ref{gfun}).

\section{\boldmath Coefficient functions $f_i(\xi)$}
\label{app:B}

The explicit expressions for the functions $f_i(\xi)$ entering the result for $\widetilde C_{\phi\phi\phi\phi}$ in (\ref{Cphi4res}) are
\begin{equation}
\begin{aligned}
   f_1(\xi) &= \xi - (1-\xi) \ln^2(1-\xi) - (1-\xi)\,\mathrm{Li}_2(\xi) + 4\xi\,g^2(\xi) , \\[1mm]
   f_2(\xi) &= - \xi + (1-\xi)\,\mathrm{Li}_2(\xi) + 2\xi\,g^2(\xi) \,, \\
   f_3(\xi) &= - \frac{5-\xi}{4} - \frac{3-4\xi+\xi^2}{4\xi} \ln(1-\xi) + \frac{1}{2\xi}\,\mathrm{Li}_2(\xi) 
    + \frac12 \sqrt{\xi(4-\xi)}\,g(\xi) - g^2(\xi) \,, \\[-1mm]
   f_4(\xi) &= - 2 + \xi - \frac{2(1-\xi)^2}{\xi} \ln(1-\xi) + 4 \sqrt{\xi(4-\xi)}\,g(\xi) - 8 g^2(\xi) \,, \\[-1mm]
   f_5(\xi) &= \xi - (1+\xi) \ln^2(1-\xi) - (1+\xi)\,\mathrm{Li}_2(\xi) - \xi\,\frac{\pi^2}{12} \,, \\[0.5mm] 
   f_6(\xi) &= - \xi + \xi\,\mathrm{Li}_2(-\xi) + (1+\xi)\,\mathrm{Li}_2(\xi) \,, \\
   f_7(\xi) &= - \frac{5-\xi}{4} - \frac{3-2\xi-\xi^2}{4\xi} \ln(1-\xi) - \frac{4-3\xi}{4\xi} \ln^2(1-\xi)
    + \frac{2+5\xi}{4\xi}\,\mathrm{Li}_2(\xi) \,, \\
   f_8(\xi) &= - 2 + \xi - \frac{2(1-\xi^2)}{\xi} \ln(1-\xi) + 2\ln^2(1-\xi) + 2\,\mathrm{Li}_2(\xi) \,, \\
   f_9(\xi) &= \frac{1-\xi}{\xi} \ln^2(1-\xi) - \mathrm{Li}_2(\xi) \,, \\
   f_{10}(\xi) &= \frac{3-2\xi}{\xi} \ln^2(1-\xi) - 3\,\mathrm{Li}_2(\xi) \,.
\end{aligned}
\end{equation}
\end{widetext}

\end{document}